\documentclass[reprint,twocolumn,amsmath,amssymb,groupedaddress,notitlepage,longbibliography]{revtex4-2}
\usepackage{graphicx,ar}
\usepackage{xcolor}

\begin{document}

\title{Thrust force is tuned by the rigidity distribution in insect-inspired flapping wings}
\author{Rom\'eo Antier} 
\author{Benjamin Thiria} 
\author{Ramiro Godoy-Diana} 
\email{ramiro@pmmh.espci.fr}
\affiliation{
Laboratoire de Physique et M\'ecanique des Milieux H\'et\'erog\`enes (PMMH), CNRS UMR 7636, ESPCI Paris--Universit\'e PSL, Sorbonne Universit\'e, Universit\'e Paris Cit\'e, F-75005 Paris, France}

\begin{abstract}
We study the aerodynamics of a flapping flexible wing with a two-vein pattern that mimics the elastic response of insect wings in a simplified manner. The experiments reveal a non-monotonic variation of the thrust force produced by the wings when the angle between the two veins is varied. An optimal configuration is consistently reached when the two veins are spaced at an angle of about 20 degrees. This value is in the range of what has been measured in the literature for several insect species. The deformation of the wings is monitored during the experiment using video recordings, which allows to pinpoint the physical mechanism behind the non-monotonic behaviour of the force curve and the optimal distribution of the vein network in terms of propulsive force.
\end{abstract}


\maketitle

\section{Introduction}

Quoting Wootton \cite{Wootton:1981}: \emph{"In considering insect wings, whether for comparative illustration or aerodynamic analysis, some simplifications are inevitable. Two in particular are common: to regard the wing as essentially flat, and as effectively rigid. Neither is true, and the latter can be seriously misleading."} And the same can be said for most flapping wings and fins, where the structural deformation that accompanies the back and forth motion is a fundamental element of the dynamical balance \cite{Shyy:2013,Hedrick:2015}. In particular, the periodic stroke reversals of flapping wings and their associated cycle of acceleration and deceleration give rise to a rich variety of vortex structures that are crucial players in the unsteady aerodynamic mechanisms inherent to flapping flight---see e.g. \cite{Sane:2003,Wang:2005,Sun:2014,Chin:2016}, for a review. Another noteworthy point is that these mechanisms are tuned with the deformation dynamics of the wings \cite{Zhao:2010,Cai:2022,Othman:2023}, where specific features such as the passive wing pitch reversal observed in some insects \cite{Bergou:2007} or the active camber control used by bats \cite{Muijres:2008} are determinant in the cycle of aerodynamic force production.  Several works \cite{Hamamoto:2007,Walker:2009,Zhao:2010} have addressed the problem of wing deformation (see e.g. \cite{Bomphrey:2018} for a review) and models have usually decomposed the main deformation modes as a combination of spanwise and chordwise bending \cite{Combes:2003_I,Combes:2003_II,Heathcote:2008}. In the case of insects, a network of veins confers their wings an anisotropic rigidity \cite{Ennos:1989,Hedrick:2015,Wootton:1992}, which governs the passive responses of the wings to aerodynamic, inertial and occasional impact forces \cite{Wootton:2020,Phan:2020}. The structural function of veins is not straightforward, and it coexists with their other roles as transmission conduits of air and hemolymph, and as sensory elements \cite{Wootton:1992,Weber:2023}. However, a few main features are recurrent, such as the veins being the thickest closest to the wing root, tapering towards the tip and trailing edge of the wings \cite{Rees:1975,Jongerius:2010}. Another observation is that most insect wings present a zone near the leading edge stiffened by thick veins and relief (see e.g. \cite{Wootton:2020}), with thinner veins elsewhere that will let the wing membrane deform more easily during the flapping motion. A secondary stiffened axis oriented obliquely at an angle with the leading edge is also present in many insect wings. Interestingly, the angle between the leading edge and this oblique stiffened area is narrowly-distributed around 15$^{\circ}$ to 30$^{\circ}$ (see e.g. for dipteran wings \cite{Ennos:1989,Tanaka:2011}); the stiffness of this zone is provided by the combined effect of veins and corrugation \cite{Tanaka:2011}. The details of the local venation are essential here, and it has been suggested that extant vein patterns have resulted from evolutionarily convergent vein fusions \cite{Cousin:2016}.

Studies on artificial wings abound in the recent literature (see e.g. \cite{Hasan:2019,Phan:2019,Wang:2023} for a review), fuelled on the one hand by the research effort on bio-inspired flapping-wing robots \cite{Wood:2008,Bao:2011,Fei:2015,Roll:2016,Zou:2016,Liu:2017,Nagai:2021}, and also by the fabrication possibilities offered by the widespread availability of 3D-printing \cite{Saito:2021}. Previous computational \cite{Vanella:2009,Nakata:2012,Nakata:2018} and experimental \cite{Thiria:2010,Ramananarivo:2011,Ryu:2019,Addo-Akoto:2021} works have investigated the relationship between wing deformation and thrust generation, showing that wing compliance (chordwise and spanwise bending and torsion) are crucial for aerodynamic performance (force production and efficiency). However, disentangling the wide variety of effects at play in the aeroelastic problem of flapping wings with a complex structural architecture is a difficult task. 
\begin{figure*}
\centering
\includegraphics[width=\textwidth]{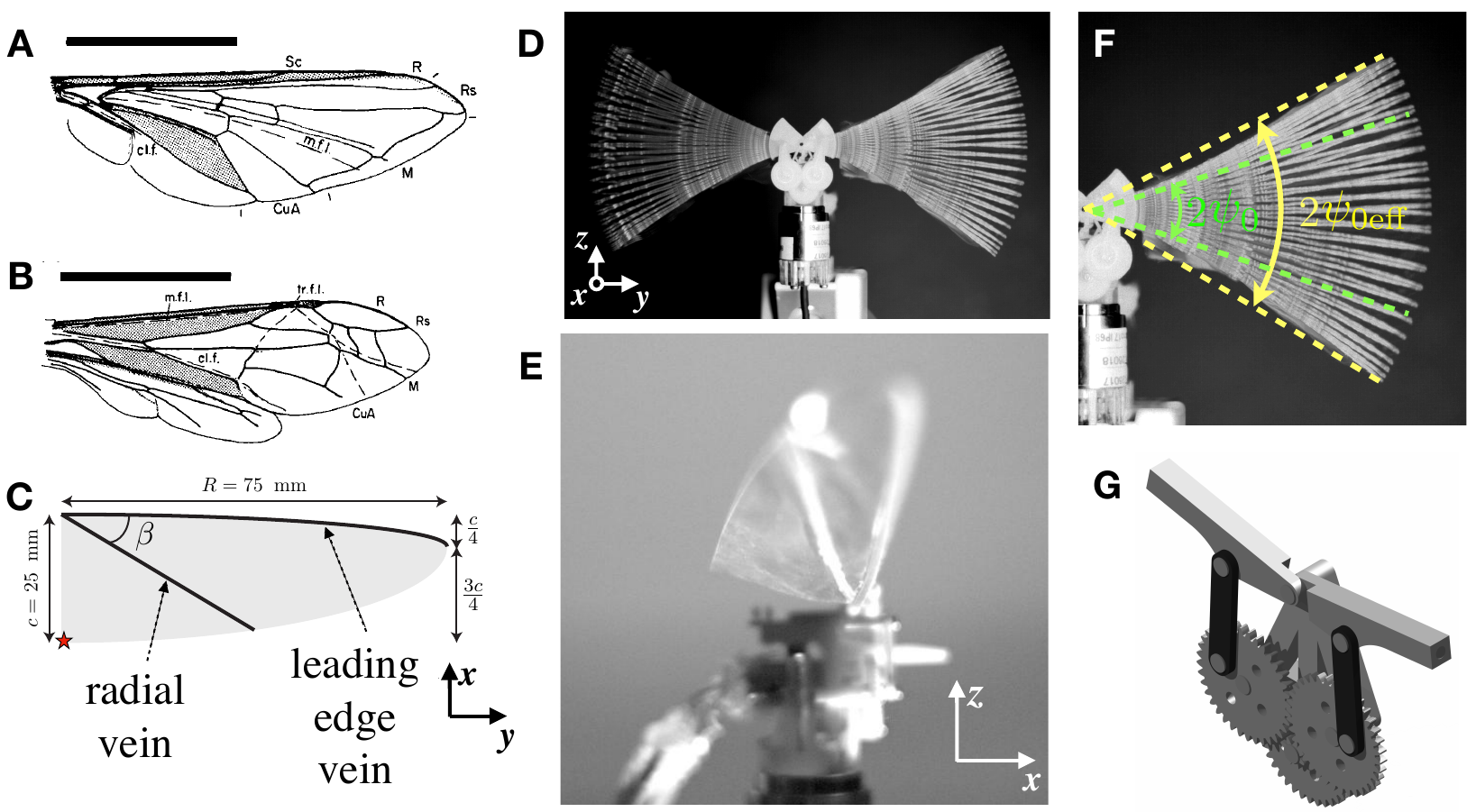}
\caption{(A) and (B) Two examples of the rigidity distribution in insect wings (Figures from \cite{Wootton:1981}). Supporting areas (stippled), deformable areas (unstippled) and flexion lines (dashed) in (A) \emph{Syrphus ribesii} (Diptera); (B) \emph{Vespula germanica} (Hymenoptera). m.f.l., median flexion line; cl.f., claval furrow; tr.f.l., transverse flexion line. Scale lines = 5 mm. (C) Model wing used in the present work. The point on the trailing edge closest to the wing root marked with a red star is the \emph{trailing edge tip}. (D) and (E) Frontal and side views, respectively, of the system mounted on the force sensor. In (D) several snapshots are superposed to illustrate the flapping wing motion. (F) Illustration of the effective stroke angle $\psi_{0\rm eff}$ compared to the imposed stroke angle at the wing base $\psi_{0}$. (G) Schematic diagram of the flapping mechanism. The electric motor is connected to the lowest gear wheel through a reduction gear.}
\label{fig:wings}
\end{figure*}

The goal of this paper is to examine the effect of different patterns of wing rigidity on the flapping wing aerodynamics by using a minimal model that mimics the elastic response of insect wings in a simplified manner.  We use a two-vein pattern (see Fig.~\ref{fig:wings}) with one main vein along the leading edge of the wing, and a secondary vein also attached at the root of the wing but that extends obliquely at a specific angle with respect to the leading edge. The angle between the two veins is the main experimental parameter explored. Our model is similar to the one investigated by \cite{Chaudhuri:2015}, but it does not have a permanent chordwise vein at the root of the wing. This allows the wing to deform more drastically when the angle between the two veins is changed.  The advantage of having a single experimental parameter (the angle $\beta$ between the leading edge vein and the radial vein) to control the deformation of the wing is that it is simple to implement. However, the present model does not accurately reproduce the torsional response of real insect wings. For example, during flapping, the wings of flies (Diptera) twist from root to tip, with the distal end of the wing twisting more than the proximal end \cite{Ellington:1984_III,Walker:2009}. This is the opposite of the torsional motion observed for most of the model wings in this study. The main reason for this difference is that insects use hinge joints at the wing attachment and muscles in the thorax to control the motion of their wings \cite{Wootton:1992}. A more complex robotic setup would be needed to mimic these control mechanisms more accurately. 
However, the main goal of this study is to examine the passive elastic response of a flexible wing with an anisotropic distribution of rigidity, using a very simple model with two rigid bars. This distribution of rigidity is inspired by the distribution of rigidity found in insects. Wing models with rigid bars can capture the flexion lines observed in real insects, which separate the different deformation areas of the wing and orient the membrane bending. Our experiments show that this bending response, which depends on the angle $\beta$ between the two veins, governs the thrust production performance of the wings.

The experiments reveal a non-monotonic variation of the average thrust force produced by the flapping wings with a local optimum when the two veins are spaced at an angle of about 20$^{\circ}$, which is in the range of the typical angles observed in insects. An explanation of the physical mechanisms involved is proposed using observations of the instantaneous kinematics of the flapping wing deformation.

\bigskip
\section{Experimental setup and methods}

\subsection{Wings and flapping system}
A flapping system with two wings is used for the experiments. The wings are composed of a 3D-printed skeleton with two veins and a thin membrane. The model wing is a Zimmerman planform (see Fig.\ref{fig:wings}~C) \cite{Wu:2010,Kang:2011}, that represents a simplified hummingbird or insect wing shape. The wings are shaped like two quarter ellipses. A quarter ellipse with half major axis the span and half minor axis one quarter of the chord. The leading edge of the wing follows the curve of this first quarter ellipse. The second quarter ellipse is connected to the first one along the span. The half major axis is also the length of the wingspan, and the half minor axis is three quarters of the chord. The length of the wing is $R = 75$ mm, and the root chord measures $c = 25$ mm, the mean chord thus being $c_m=19.6$ mm. The aspect ratio of the wing is $\AR=R/c_m = 3.82$. The wing shape is cut on a polyethylene terephthalate (PET) film of thickness 30~$\mu$m. The cutting is done with a laser cutting machine (CO2 with infrared ray, Epilog Laser, type Helix). The Young's modulus of the membrane is $4 \pm 0.3$ GPa, measured with 3 experiments of tensile test. We used a tensile test machine Instron 5865 mounted with a static load cell of maximum capacity 1 kN, and we imposed the deformation of the samples by fixing a constant displacement velocity of 1 mm/min. Samples dimensions are 100 mm $\times$ 40 mm $\times$ 30 $\mu$m.

\begin{table*} 
\caption{\fontsize{9}{9}\selectfont Wing morphological and material properties, kinematic parameters, and dimensionless numbers for a few insect species and for the model wings. $m_b$ is the total body mass of the insects, $m_w$ is one wing mass used to compute the average surface mass density $\mu_s=m_w/S$. Data from \cite{Magnan:1934,Greenewalt:1962,Dudley:1990,Combes:2003_I,Shyy:2013}.}
\label{dimensionless_numbers}
\setlength\tabcolsep{0pt} 
\smallskip 
\begin{tabular*}{\textwidth}{@{\extracolsep{\fill}}llccccc}
\hline
Parameter &units &Hawkmoth  & Hoverfly  & Bumblebee & European honey bee  & Model wings \\
& & \emph{(Manduca sexta)} & \emph{ (Eristalis tenax)} &  \emph{(Bombus terrestris)} 
& \emph{(Apis mellifera)} & of the present study   \\      
\hline
\\
$R$ &(m) &0.049  & 0.009 & 0.016 & 0.0097 & 0.075\\ 
$S$ &(m$^2$) &8.91$\times 10^{-4}$ & 3.7$\times 10^{-5}$ & 1.1$\times 10^{-4}$ & 4.2$\times 10^{-5}$ & 0.0015\\ 
$c_m$ &(m) &0.018 & 0.004 & 0.007 & 0.004 & 0.02\\ 
$m_b$ &(kg) & $1.6\times 10^{-3}$ & $0.09\times 10^{-3}$ & $0.18\times 10^{-3}$& $0.1\times 10^{-3}$& - \\	
$m_w$ &(kg) &4.7$\times 10^{-5}$ & 6.0$\times 10^{-7}$ & 1.25$\times 10^{-6}$ & 5.0$\times 10^{-7}$ & 1.5$\times 10^{-4}$\\ 
$\mu_s$ &(kg m$^{-2})$ &0.0527 & 0.0162 & 0.0118 & 0.0119 & 0.1018\\ 
$^\dag EI_{\rm beam}$ &(N m$^2$) &8.0$\times 10^{-6}$ & 7.7$\times 10^{-6}$ & 7.7$\times 10^{-6}$ & 1.82$\times 10^{-6}$ & [0.77 - 47.5]$\times 10^{-6}$\\ 
$^\dag\overline{EI}$ &(N m) &1.63$\times 10^{-4}$ & 8.56$\times 10^{-4}$ & 4.81$\times 10^{-4}$ & 1.88$\times 10^{-4}$ & [1 - 63]$\times 10^{-5}$\\ 
\\
$\psi_0$ &($^{\circ}$; rad) &57; 0.99 & 51; 0.90  & 60; 1.05 & 65; 1.13 & $^\ddag$28; 0.49\\ 
$f$ &(Hz)&25 & 210 & 150 & 250 & [12 - 20]\\ 
\\
$Re$& -&2933 & 934 & 2204 & 1577 & [1144 - 1906]\\ 
$C_Y$& -&0.131 & 0.001 & 0.009 & 0.008 & [0.006 - 0.952]\\ 
$\mathcal{N}_{ei}$& -&2.34 & 0.02 & 0.11 & 0.14 & [0.25 - 43.49]\\ 
$k$& -&1.17 & 1.58 & 1.24 & 1.23 & 1.68\\
$\AR$&- &2.69 & 3.10 & 2.18 & 2.41 & 3.82\\ 
\hline
\end{tabular*}
\footnotesize
\begin{tabular*}{\textwidth}{@{\extracolsep{\fill}}l}
$^\dag$the values $\overline{EI}$ of average plate bending rigidity  were obtained as $\overline{EI}=EI_{\rm beam}/R$, where $EI_{\rm beam}$ is the chord-wise flexural stiffness.\\ $EI_{\rm beam}$ values for insects \cite{Combes:2003_I} come from an indirect measurement of an equivalent beam performed by applying a point force to bend \\ the wing and using the measured force $F$ and wing displacement $\delta$ to calculate an overall flexural stiffness $EI_{\rm beam}=Fl^3/3\delta$, $l$ being the\\ effective beam length. $^\ddag$For the model wings the value used for the stroke amplitude is $\psi_{0\rm eff}$ (see Fig.\ref{fig:wings}~F). 
\end{tabular*}
\end{table*}%

The membrane is supported by two 3D-printed reinforcements disposed respectively on the leading edge and along a direction making an angle $\beta$ with the leading edge (see Fig.\ref{fig:wings}~C). The reinforcements play the role of veins. The material of the veins is polylactic acid (PLA) with a Young's modulus of $2.35$ GPa (green 3D printing PLA filament sold by Ultimaker). The 3D printer used was an Artillery Sidewinder X2.  In order to have a symmetric deformation, a vein skeleton is glued on each side of the membrane using a Teroson SB2444 rubber adhesive. This adhesive is very elastic and allows the membrane to slide between the veins without detaching. The veins are 1 mm wide. The thicknesses on each side of the membrane are of 480 $\mu$m for the leading edge and 240 $\mu$m for the radial vein. The angle between the two veins $\beta$, varies between 10$^{\circ}$ and 90$^{\circ}$ in 5$^{\circ}$ steps. We also made a wing consisting of a membrane and a single vein at the leading edge. This wing is referenced by the case $\beta=0^{\circ}$. The leading edge vein is extended by 3.5 mm toward the center of the ellipse so that it can be connected to the flapping mechanism. The length of the radial vein, $R_v$, evolves as a function of the angle $\beta$ through the relationship: 

$$R_v = \frac{1}{\cos(\beta)} \left(\frac{1}{R^2} + \frac{\tan^2\beta}{c^2} \right)^{-1/2}$$

Because $R_v$ diminishes with increasing $\beta$, the total mass of the wing also diminishes slightly. The difference in mass between the heaviest wing ($\beta=5^{\circ}$) and the lightest wing ($\beta=90^{\circ}$) is 16\%.

Two wings are mounted on a flapping system based on the DelFly design \cite{deCroon:2016} obtained by dismantling a commercially-available flapping-wing bird toy (Dilwe RC Flying Bird, ASIN B09J2NWSVK) to keep only the motor and crank mechanism ---see Fig.~\ref{fig:wings} (G). The system, powered externally, allows to generate a sinusoidal planar flapping motion with a stroke amplitude of $2\psi_0=32^{\circ}$ ---see Fig.~\ref{fig:wings} (F)--- for frequencies $f$ ranging from 5 to 20 Hz. Front view and side view photos of the system are shown in Fig.~\ref{fig:wings} (D) and (E), respectively.

\subsection{Wings scaling analysis}

In order to asses how far, dynamically, our rudimentary model wings are from the case of an insect wing, it is convenient to examine a few dimensionless quantities. The main elements of the flapping wing problem are the aerodynamic force, the elastic bending rigidity of the wing, and its inertia \cite{Daniel:2002,Thiria:2010,Wu:2019}. We can use as a basic model a flexible plate (dimensions: mean chord $c_m$, span $R$ and thickness $h$, density $\rho_s$ ---i.e. surface mass density $\mu_s=\rho_sh$---, and elastic modulus $E$) that will bend under the action of its own inertia and of the aerodynamic forces. For a flapping motion in hovering characterised by an angular amplitude $\psi_0$ and frequency $\omega=2\pi f$, before considering the wing deformation we can already define \cite{Shyy:2013}: (i) the Reynolds number, written in terms of a reference flapping velocity  $U_{\rm ref}=2\psi_0 Rf$, the density $\rho$ and dynamic viscosity $\eta$ of air, and using the mean chord $c_m$ as reference length scale $L_{\rm ref}$:
\begin{equation}\label{eq:Re}
Re=\frac{\rho U_{\rm ref}L_{\rm ref}}{\eta}=\frac{\rho 2\psi_0 Rf c_m}{\eta} \;,
\end{equation}

\noindent which governs the aerodynamic regime by setting the importance of fluid inertial versus viscous effects; and (ii), the reduced frequency
\begin{equation}\label{eq:k}
k=\frac{\omega c_m}{U_{\rm ref}}=\frac{\pi}{\psi_0 \AR} \;,
\end{equation}

\noindent which in the present hovering case does not depend explicitly on the physical frequency because the reference velocity is the flapping velocity that is itself proportional to the frequency.

Now, to estimate the effects of aerodynamic loading and wing inertia measured against the elastic response of the wing, we can use, respectively, a Cauchy number \cite{deLangre:2008,Ishihara:2009}:

\begin{equation}\label{eq:Cy}
C_Y = \frac{\frac{1}{2} \rho U_{\rm ref}^2 L_{\rm ref}^3}{\overline{EI}} = \frac{2\rho R^2 \psi_0^2 f^2 c_m^3}{\overline{EI}} \;,
\end{equation}

\noindent which characterizes the deformation of the wing under the effect of the fluid flow, and the elasto-inertial number \cite{Thiria:2010}:

\begin{equation}\label{eq:Nei}
\mathcal{N}_{ei}=\frac{\mu_s a_{\rm ref}L_{\rm ref}^3}{\overline{EI}}=\frac{4\pi^2\mu_sR\psi_0f^2c_m^3}{\overline{EI}} \;,
\end{equation}

\noindent which characterizes the deformation of the wing under the effect of its own inertia. $\mathcal{N}_{ei}$ is written in terms of a reference acceleration $a_{\rm ref}=R\psi_0\omega^2$. $\overline{EI}$ in Eqs.~\ref{eq:Cy} and \ref{eq:Nei}  is an average plate chordwise bending rigidity. We consider the chordwise bending rigidity because it is the one that governs the transverse deformation of the wings. Its value changes dramatically with $\beta$ because of the effect of the radial vein. The range of values representing all wings are reported in Table \ref{dimensionless_numbers}. The bending rigidity was measured by bending tests as in \cite{Combes:2003_I}. A brief summary of the method is recalled in a footnote to Table \ref{dimensionless_numbers}. A displacement was imposed on different locations on the veins and the associated force was measured. We measured the rigidity on veins only because it was technically impracticable to measure something by applying a load on the membrane. We thus measured the bending rigidity spanwise (mostly dictated by the bending of the leading-edge vein) and chordwise (although we could call it radial-vein-wise). We observe that the spanwise rigidity is independent of $\beta$ whereas the chordwise rigidity is very dependent on $\beta$. This mechanism explains the wide range of measured values. In Table \ref{dimensionless_numbers}, we consider the chordwise rigidity because it is the one that governs the transverse deformation of the wings. The spanwise bending rigidity governs the deformation of the leading edge and is responsible for the effective stroke angular amplitude $\psi_{0\rm eff}$ being greater than the imposed value $\psi_0$, but it is the same for all wings.

These dimensionless numbers can be used to give an indicative picture of the model wings in comparison with insect (or other) wings in a global parameter space, as shown in Table \ref{dimensionless_numbers}, and comfort the idea of using the present experiment to examine the effect of the rigidity distribution of the wing on its aerodynamic performance. 

\begin{figure*}
\centering
\includegraphics[width=0.85\linewidth]{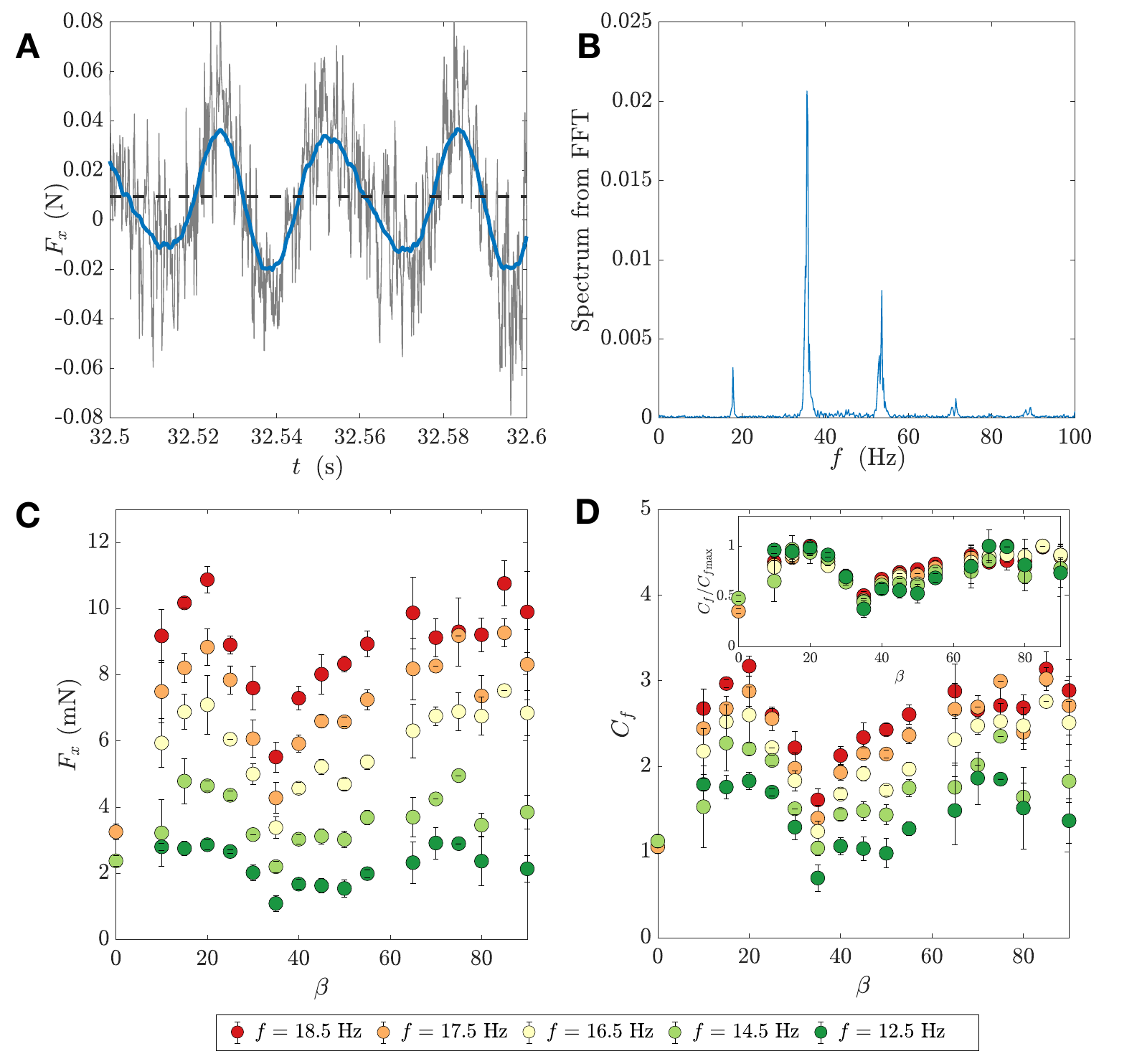}
\caption{Thrust force measurements. (A) Time series $F_x(t)$ and (B) its frequency content obtained by FFT for a typical case ($\beta= 20^{\circ}$ and $f= 17.5$ Hz). In (A) a running mean of the signal is shown (solid blue line) as well as its average value (dashed line). (C) Average force $\bar F_x$ as a function of the radial vein angle $\beta$ for several frequencies ($f=12.5$, 14.5, 16.5, 17.5 and 18.5 Hz, from darkest to lightest color respectively). Each point is the mean value of several runs with identical parameters and the corresponding standard deviations are represented as error bars. (D) Thrust coefficient $C_f$ for the same experimental data computed using Eq.~\ref{thrust_coeff}; in the inset, $C_f$ is normalized by the maximum value $C_{f{\rm max}}$ of each frequency series. }
\label{fig:force}
\end{figure*}

\subsection{Force sensor}

The system is mounted on a Schunk FT-Nano 17 6-axis force sensor as shown in Fig.~\ref{fig:wings}~D, such that the average propulsive force produced by the flapping wings points towards the $x$-direction. In a right-handed cartesian reference frame, the weight of the device is thus directed towards the negative $z$-direction. In what follows, we focus on the forward component of the force $F_x$, which in the present setup is the most relevant concerning the aerodynamic force production because the forward component is perpendicular to the stroke plane (as in the merry-go-round setup of \cite{Thiria:2010, Ramananarivo:2011}). The reciprocal motion of the wings and the symmetry of the setup determine that the $F_y$ and $F_z$ components of the force as well as the $M_x$ and $M_z$ components of the moment average to zero over each flapping period. The $M_y$ component has non-zero mean, but what can be learned from its dynamics in the tethered frame of the present experiment is redundant from what is obtained from the analysis of the forward force $F_x$.

A typical time series of the measured force signal is shown in Fig.~\ref{fig:force} (A). The signal is noisy because the forces produced by the wings were close to the limit of the measurement range of our sensor. Nonetheless, the periodicity driven by the flapping motion is clearly visible, as highlighted in the figure by the running average also plotted. These time-resolved measurements were robustly repeatable. The Fourier transform of the signal---shown in Fig.~\ref{fig:force} (B)---has its largest peak at twice the flapping frequency. This is expected because two peaks of force are produced over one period when the wing instantaneous velocity is highest during the upstroke and the downstroke. The average force, marked as a dashed horizontal line in Fig.~\ref{fig:force} (A), is the main output used as a performance probe as the $(\beta, f)$-parameter space is explored. The results are summarised in Fig.~\ref{fig:force} (C), where this time-averaged force $\bar F_x$ is plotted as a function of the radial vein angle $\beta$ for several frequencies. For each point, at least 4 experiments were conducted: 2 experiments with one pair of wings and 2 with a second identical pair of wings. Each force is averaged over a 5-second steady-state period, this period corresponds to at least 50 flapping cycles. We may note that the magnitude of the forces produced by the model wings are small compared to the weight of an insect of comparable wing size. The present system has thus no chance of taking off, but it will serve its purpose of examining the physical mechanisms that relate changes in the stiffness distribution of the wings to the production of aerodynamic force.

\begin{figure*}
\centering
\includegraphics[width=0.99\textwidth]{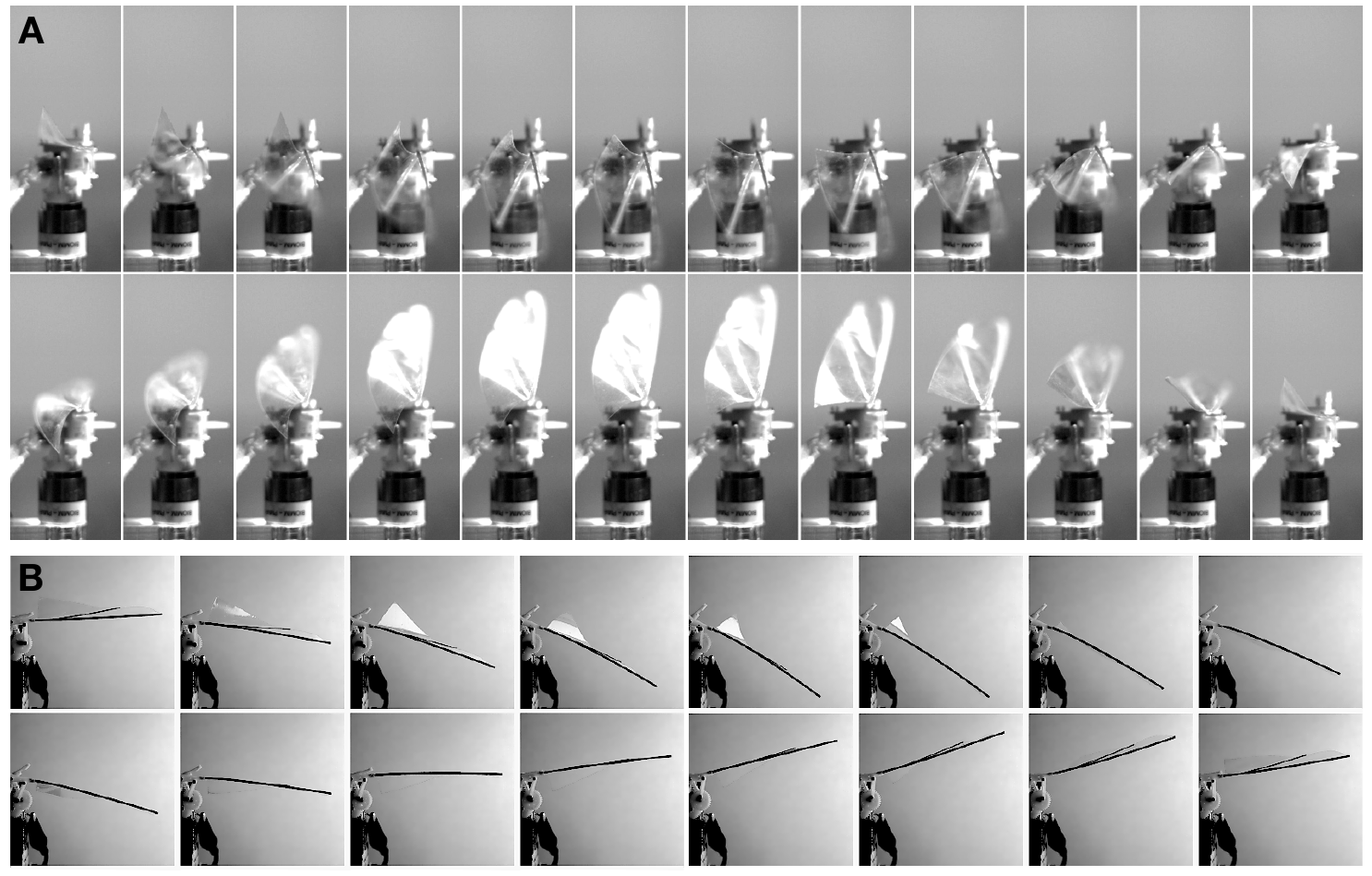}
\caption{(A) Side and (B) front views of a flapping sequence of a wing with $\beta=40^{\circ}$ at 17.5 Hz. One flapping cycle is represented starting from the $\psi=0$ position at mid downstroke. The time interval between snapshots is of 2.38ms in (A) and 3.57ms in (B).}
\label{fig:kinematics}
\end{figure*}
\subsection{Kinematics tracking}

In addition to the force measurements, the motion of the wings was tracked using a Phantom Miro M120 high-speed camera recording $1920\times 1200$ pixel$^2$ images at 800Hz. Fig.~\ref{fig:kinematics} shows time series of a side view (A) and a front view (B) to give a qualitative picture of the deformation of the wing during the flapping cycle. In order to quantify the wing deformation, four points of interest were tracked using ImageJ \cite{ImageJ}: two at the leading edge (at the root and at the tip), and two at the trailing edge (at the point where the radial vein ends, and at end of the largest chord section). 

\begin{figure*}
\centering
\includegraphics[width=0.89\linewidth]{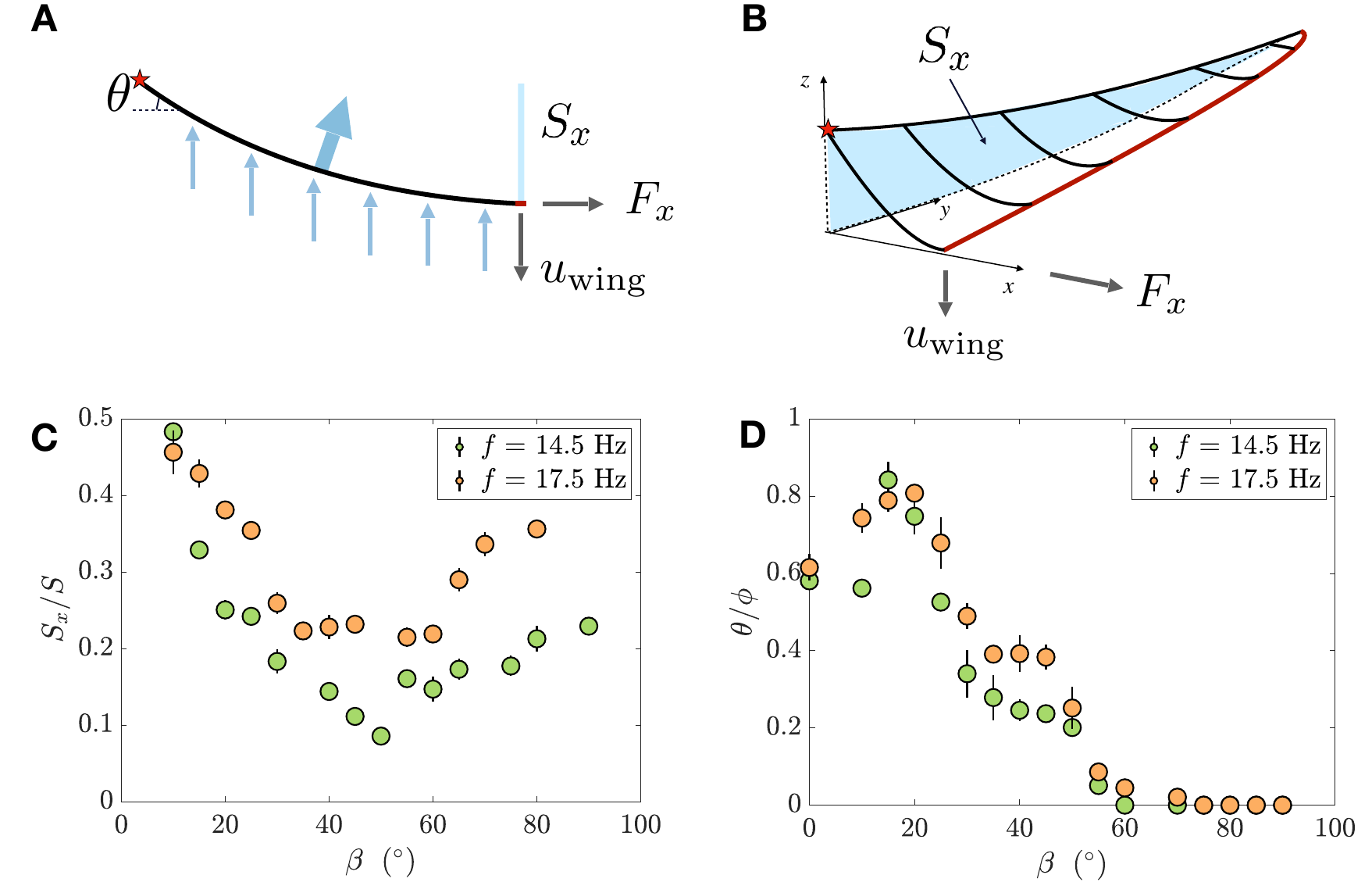}
\caption{(A) and (B) Schematic representations of the flexible wing moving at speed $u_{\rm wing}$. In (A), a section in the $xz$ plane is pictured, the thick blue arrow represents the total aerodynamic force which points more in the forward direction the more the wing is bent. The trailing angle $\theta$ and the projected surface $S_x$ are also shown. (B) shows a three-dimensional sketch. The trailing edge tip marked with a red star is the point tracked to measure $\theta$. (C) and (D) Measured values of the projected surface $S_x$ (C) and the trailing angle $\theta$ (D) as a function of $\beta$ for $f=14.5$ and 17.5 Hz.}
\label{fig:sketch_wing}
\end{figure*}

\section{Results and discussion}

\subsection{Thrust coefficient}

As mentioned above, the time average of the force $\bar F_x$  presented in Fig.~\ref{fig:force} (C) is the main performance indicator of the system as a function of the parameter space constituted by the radial vein angle $\beta$ and the flapping frequency $f$. A first step in the analysis is naturally to find a dimensionless representation of the data that is plotted in physical units in Fig.~\ref{fig:force}. To do so, we define a thrust coefficient

\begin{equation}
C_f=\frac{\bar F_x}{\frac 1 2 \rho u_{\rm wing}^2 S}
\label{thrust_coeff}
\end{equation}

\noindent where  $u_{\rm wing}=2\pi f A$ is the characteristic flapping speed defined by the frequency $f$ and amplitude $A$ of the flapping motion. We define a nominal amplitude $A=R\sin(32^{\circ})$ based on the wing length and the flapping angular amplitude. The reference surface $S$ is the area of the two wings. The thrust coefficient is presented in Fig.~\ref{fig:force}~(D). Two main observations can be made: on the one hand, the measurements for each frequency constitute a non-monotonic curve. As the angle $\beta$ of the radial vein is increased from zero, a  clear maximum occurs around $\beta\approx 20^{\circ}$, followed by a minimum at $\beta\approx 35^{\circ}$. Further increasing $\beta$ makes the propulsive force grow again until it reaches a similar value to that of the first maximum observed at $\beta\approx 20^{\circ}$. The second observation is that when the frequency is increased, the performance curve is shifted to higher values whilst keeping a fairly similar shape. This can be seen clearly by normalising the curve corresponding to each flapping frequency by its maximum value $C_{f{\rm max}}$, as shown in the inset of Fig.~\ref{fig:force}~(D). That the thrust force increases with the flapping frequency is of course an expected result, which has been reported for similar systems in the literature \cite{Mazaheri:2010,Nguyen:2016}.

In what follows we analyse these results in light of the wing deformation observations.

\subsection{Wing deformation kinematics}

An overview of the flapping motion captured from side and top views is presented in Fig.~\ref{fig:kinematics}. The main point of this visualisation is to examine the typical behaviour of the wing and to identify the basic elements of its deformation dynamics. The phase lag between the leading edge and trailing edge has been used in the literature \cite{Ramananarivo:2011} to explain the performance increase of flexible wings with respect to rigid wings. Considering that the average main component of the aerodynamic force is perpendicular to the stroke plane, the advantage of the flexible wing comes from the redirection of the force in the useful direction. A representation of this idea for a section of a flexible wing is shown in Fig~\ref{fig:sketch_wing} (A), which defines the two parameters that will be used in the following:  the projection of the wing surface on the stroke plane $yz$, defined as $S_x$ ---see also Fig~\ref{fig:sketch_wing} (B)---, and the trailing angle $\theta$, both measured at the instant of maximum flapping velocity (i.e. when the wing passes the horizontal position). Note that the trailing angle $\theta$ is different from the usual aircraft trailing edge angle, defined as the angle between the tangents of the upper and lower airfoil at trailing edge, indicating trailing edge sharpness. Now, $S_x$ can be used as a measure of the aforementioned force redirection. If the wing is considered as a homogeneous plate bending under its first mode of deformation, this picture is sufficient to explain the basic physical mechanism driving the performance of a flexible wing \cite{Thiria:2010,Zhao:2010}. The radial vein complicates the picture because each section of the wing behaves now differently ---see Fig.~\ref{fig:wing_sections} (B) and (C). In particular, the phase lag of the trailing edge becomes different depending on the span-wise position. Nonetheless, we can still examine how $S_x$ changes with $\beta$. This is shown in Fig.~\ref{fig:sketch_wing} (B) for the cases of $f=14.5$ and 17.5 Hz. As $\beta$ increases from zero, the radial vein starts preventing part of the wing to bend and $S_x$ diminishes. This trend saturates at $\beta\approx 40^{\circ}$ until  $\approx60^{\circ}$, after what $S_x$  increases again to larger values. Recalling the force measurements of Fig.~\ref{fig:force}, where the thrust minima are observed  around $\beta\approx 40^{\circ}$, reinforces the idea of a larger projected surface $S_x$ being a necessary feature for increasing thrust production.

\subsection{Projected wing surface and trailing angle}

\begin{figure}[t]
\centering
\includegraphics[width=\linewidth]{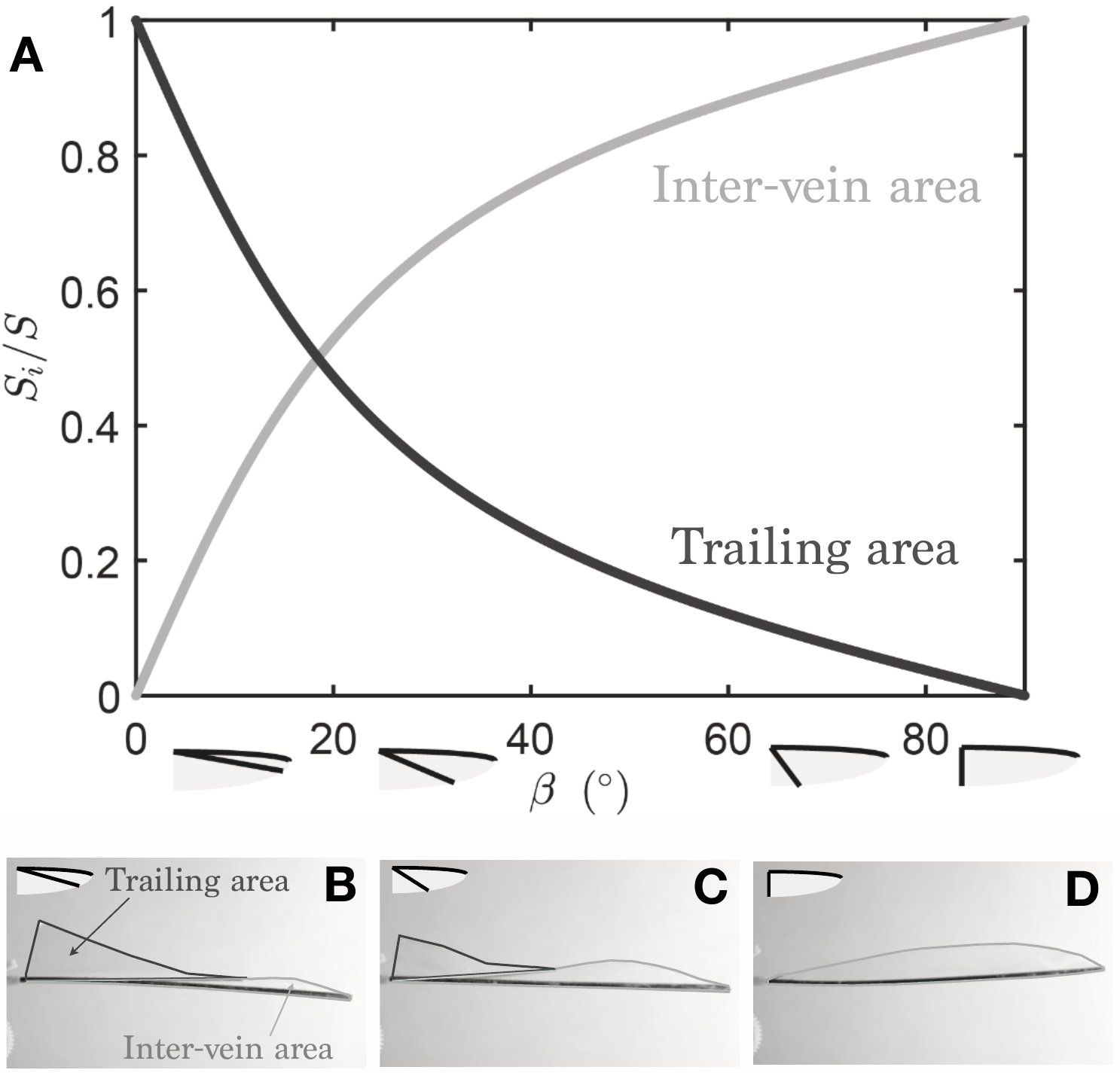}
\caption{(A) Areas of the inter-vein and trailing sections of the wing as a function of $\beta$; (B), (C), and (D) frontal snapshots of the wing at $\psi=0$ for different values of $\beta$ with the wing sections highlighted.}
\label{fig:wing_sections}
\end{figure}

To go further, it is useful to describe the wing in terms of its two sections: the first section is the area comprised between the leading edge and the radial vein and the second one that between the radial vein and the trailing edge. We define these, respectively, as the \emph{inter-vein area} and the \emph{trailing area}. Because the wing is built with portions of ellipses, the areas of these two sections can be expressed analytically, as a function of $\beta$, which gives the curves shown in Fig.~\ref{fig:wing_sections} (A). Panels (B), (C), and (D) in Fig.~\ref{fig:wing_sections} show three examples for different angles of the radial vein, when the wing passes the horizontal position, with the perimeters of the two areas highlighted. Since the snapshots are frontal views of the wing, the addition of the highlighted areas is actually $S_x$. 

For small angles between the veins, typically $\beta<20^{\circ}$, the trailing area is larger than the inter-vein area, this implies that $S_x$ is dominated by the deformation of the trailing area ---see Fig.~\ref{fig:wing_sections} (B). The larger this free surface, the larger its deformation. On the contrary, for large $\beta$, typically $\beta>50^{\circ}$, the trailing area is very small and hardly deforms at all. Its influence is then small in the generation of aerodynamic forces. The surface between the veins is the largest and its swelling is at the origin of the redistribution of the aerodynamic forces ---see Fig.~\ref{fig:wing_sections} (D) for the limit case of $\beta=90^{\circ}$ where the trailing area has vanished and the whole wing is the inter-vein area. 
\begin{figure*}
\centering
\includegraphics[width=0.95\linewidth]{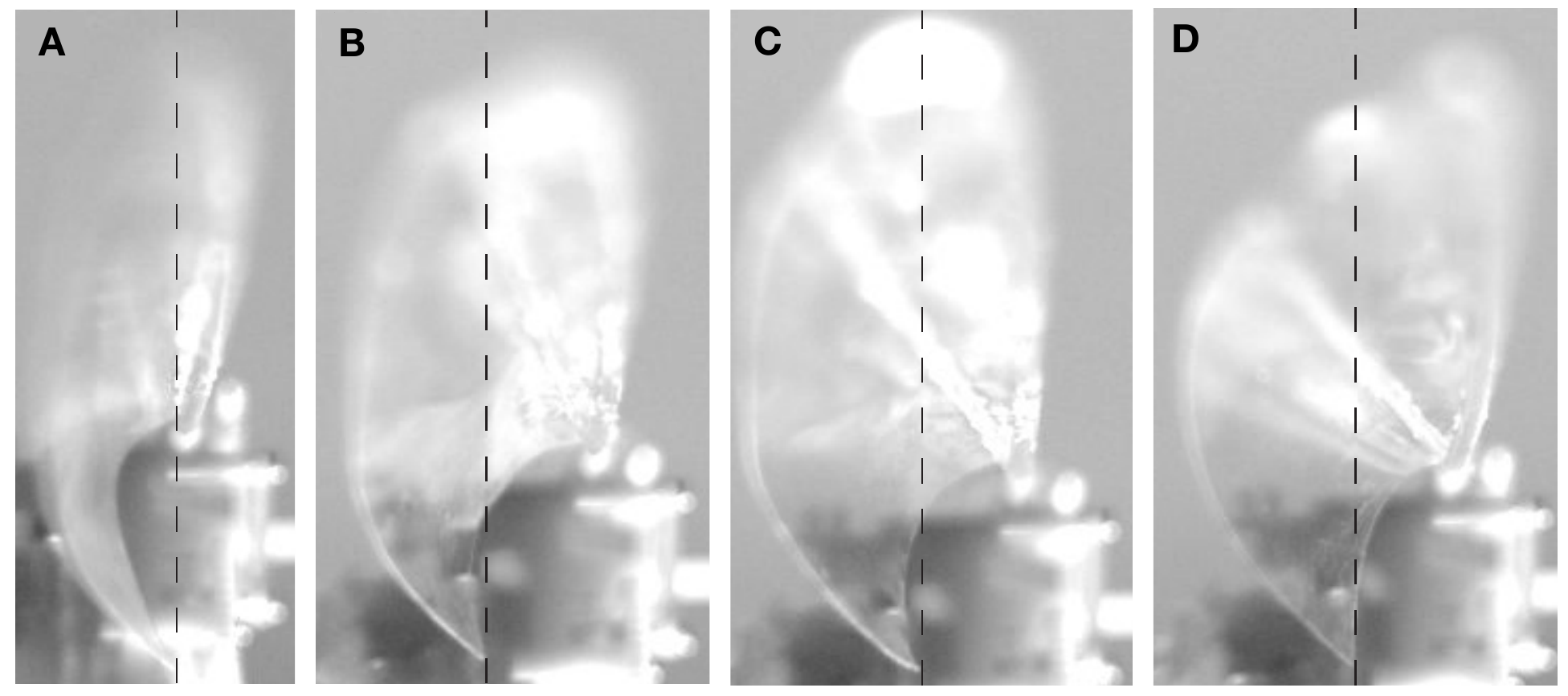}
\caption{Side view snapshots showing the trailing area bending for wings beating at 17.5Hz. (A) $\beta=0^{\circ}$. (B) $\beta=10^{\circ}$. (C) $\beta=15^{\circ}$. (D) $\beta=20^{\circ}$. The wing with $\beta=20^{\circ}$ does not bend on itself at no instant in the flapping cycle.}
\label{fig:wing_bend_side_view}
\end{figure*}
Summarising, the changes in the projected area as a function of the radial vein angle give us a first physical insight to explain the non-monotonic behaviour of the propulsive force observed in Fig~\ref{fig:force}. For lower values of $\beta$, the force production is dominated by the trailing area, whereas for higher values, typically above $\beta\approx 40^{\circ}$, it is the inter-vein area that contributes the most. 

Now, considering firstly the lower angles, say $\beta\lesssim40^{\circ}$, the force measurements have a maximum value at $\beta\approx 20^{\circ}$. This means that the thrust force does not solely depend on the surface (or projected surface) of the trailing area, which is largest at $\beta=0^{\circ}$, but also on the way the wing is bent relatively to the incident wind. Examining side views of the flapping wing (see Fig.~\ref{fig:wing_bend_side_view}) brings evidence of the reason for the suboptimal performance of the wings with very low values of $\beta$. Because the radial vein also imposes the maximum length of the trailing area that can bend, a small angle $\beta$ means that the trailing surface is long enough to bend on itself as shown in Fig.~\ref{fig:wing_bend_side_view}, thus losing the aerodynamic benefit of flexibility. In practice, the excessive bending of the trailing area determines that its orientation is suboptimal during large portions of the flapping cycle. 

This can be examined quantitatively by tracking the trailing angle---see Fig.~\ref{fig:sketch_wing} (A)---at the instant of maximum flapping velocity, as presented in Fig.~\ref{fig:sketch_wing} (D) as a function of $\beta$ for two different frequencies. We use the dimensionless representation $\theta/\phi$ introduced by \cite{Ramananarivo:2011}, where the angle of the incoming wind $\phi$ is in the present case equal to $90^{\circ}$ since there is no incident velocity on the static wing because the system is fixed in the lab reference frame. The measurements of the trailing angle in Fig.~\ref{fig:sketch_wing} (D) bring a clear explanation underlying the maximum of aerodynamic force measured for the wing with the radial vein at $\beta\approx 20^{\circ}$: as in \cite{Ramananarivo:2011}, this optimum coincides with the best alignement of the trailing angle $\theta$ and the angle $\phi$ of the local wind seen by the translating wing. As $\beta$ increases, the trailing angle goes to zero because it is measured at the tip, which is part of the trailing area that deforms less and less and behaves as in a rigid wing when $\beta$ tends to $90^{\circ}$. For these larger values of $\beta$ the influence of the trailing area concerning thrust production diminishes, so the measurement of the trailing edge as in Fig.~\ref{fig:sketch_wing} (D) becomes irrelevant. As $\beta$ becomes larger, the main part of the force production is ensured by the inter-vein area, which represents most of the total wing surface. 

Coming back to  Fig.~\ref{fig:sketch_wing} (C), the increase of the projected area $S_x$ for $\beta>50^{\circ}$ is driven by the swelling of the inter-vein area, which can be seen in Fig.~\ref{fig:wing_sections} (D) for the limit case of $\beta=90^{\circ}$. We can hypothesise that the physical mechanism enhancing thrust at these large values of the radial vein angle $\beta$ should be similar to the case described for the trailing area. However, a quantitative picture would need the tracking of the whole trailing edge and not just of a single point as we have done to produce Fig.~\ref{fig:sketch_wing} (D), which is out of reach of the experiments reported here.

\section{Concluding remarks}

We have examined the role of non-homogeneous wing stiffness in the aerodynamic force production by flapping wings, using a simple model with a two-vein skeleton. The main observation is that the thrust produced by the wings varies non-monotonically with changes in the angle $\beta$ between the two veins that constitute the skeleton: the leading edge vein and the radial vein. A local optimum of the aerodynamic performance is observed for $\beta\approx 20^{\circ}$, which is compatible with the typical angles observed in several insect wings \cite{Ennos:1989}. The radial vein in the model used here is of course a crude simplification of the complex patterns found in real insect wings, but it serves the purpose of separating the wing surface in two areas that have a dynamic equivalence to what is observed in nature. What we have called the inter-vein area functions in a similar manner to the part of the wing close to the leading edge in insects, which is rather stiff, while the trailing area deforms much more during the flapping cycle. Coupling thrust force measurements with visualisation of the wing deformation lets us explain the physical origin of the non-monotonic behaviour of the aerodynamic force production: increasing the radial vein angle makes the wing change from a regime dominated by the trailing area, with its associated strong deformations, to another regime where the inter-vein area pilots everything. It is the former case with lower angles that allows us to come back to the case of insects mentioned above: the observed optimum angle $\beta\approx 20^{\circ}$ constitutes a trade-off between using the aerodynamic benefit of deformation that redirects the average force to have a stronger thrust component, and avoiding an excessive folding of the flexible wing that diminishes its effective surface. A word of caution should be said about the limitations of the present artificial flapping wings to represent the far more complex cases of real insect wings. A first point concerns the simple up-and-down movement of the wings used here, which does not involve any of the mechanisms insects use to control wing kinematics through their thoracic muscles and hinge joints \cite{Wootton:1992}. One major feature that is thus missing is the wing rotation that accompanies flapping. In addition, the two-vein wing design does not tightly control the wing shape. This leads, firstly, to camber profiles that are not representative of real wings (see, e.g., \cite{Walker:2009} for the case of hoverflies), and, secondly, to an exaggerated lack of constrain to bending of the trailing area. These issues should be considered in the design of future insect-inspired flapping robots. Ongoing work is concerned with the study of three-dimensional wing deformation dynamics with simultaneous measurement of aerodynamic forces.

\acknowledgements{This work was supported by the Agence Nationale de la Recherche and the ASTRID program through the project ANR-19-ASTR-002 (NANOFLY).}


%

\end{document}